\documentclass{emulateapj}

\slugcomment{To appear in the {\em Astrophysical Journal}}
\shorttitle{Low-Mass Active Galactic Nucleus GH\,10}
\shortauthors{Wrobel et al.}
\begin{document}
\title{Steep-Spectrum Radio Emission from the Low-Mass 
       Active Galactic Nucleus GH\,10}
\author{J. M. Wrobel\altaffilmark{1},
        J. E. Greene\altaffilmark{2},
        L. C. Ho\altaffilmark{3},
    and J. S. Ulvestad\altaffilmark{1}}

\altaffiltext{1}{National Radio Astronomy Observatory, P.O. Box O,
Socorro, NM 87801; jwrobel@nrao.edu, julvesta@nrao.edu}
\altaffiltext{2}{Department of Astrophysical Sciences, Princeton
University, Princeton, NJ 08544; jgreene@astro.princeton.edu}
\altaffiltext{3}{The Observatories of the Carnegie Institution of
Washington, 813 Santa Barbara Street, Pasadena, CA 91101;
lho@ociw.edu}

\begin{abstract}
GH\,10 is a broad-lined active galactic nucleus (AGN) energized by a
black hole of mass 800,000~$M_\odot$.  It was the only object detected
by Greene et al.\ in their Very Large Array (VLA) survey of 19
low-mass AGNs discovered by Greene \& Ho.  New VLA imaging at 1.4,
4.9, and 8.5~GHz reveals that GH\,10's emission has an extent of less
than 320~pc, has an optically-thin synchrotron spectrum with a
spectral index $\alpha = -0.76\pm0.05$ ($S_\nu \propto
\nu^{+\alpha}$), is less than 11\% linearly polarized, and is steady -
although poorly sampled - on timescales of weeks and years.
Circumnuclear star formation cannot dominate the radio emission,
because the high inferred star formation rate, 18~$M_\odot$~yr$^{-1}$,
is inconsistent with the rate of less than 2~$M_\odot$~yr$^{-1}$
derived from narrow H$\alpha$ and [O~II] $\lambda$3727 emission.
Instead, the radio emission must be mainly energized by the low-mass
black hole.  GH\,10's radio properties match those of the
steep-spectrum cores of Palomar Seyfert galaxies, suggesting that,
like those Seyferts, the emission is outflow-driven.  Because GH\,10
is radiating close to its Eddington limit, it may be a local analog of
the starting conditions, or seeds, for supermassive black holes.
Future imaging of GH\,10 at higher linear resolution thus offers an
opportunity to study the relative roles of radiative versus kinetic
feedback during black-hole growth.
\end{abstract}


\section{Motivation}

Although little is known about the mass function of nuclear black
holes with mass below about $10^6~M_\odot$, this is an important
regime for several reasons.  For example, the remarkable correlation
between host bulge velocity dispersion and black hole mass strongly
hints that black holes play an essential role in galaxy evolution
\citep{kor04}, yet there are few observational constraints on the
starting conditions, or so-called seeds, for the supermassive black
holes with masses above about $10^6~M_\odot$.  Also, until the {\em
Laser Interferometer Space Antenna\/} launches, an understanding of
the low-mass end of the black-hole mass function may provide one of
the few constraints on these seed black holes \citep{hug02}.
Moreover, it remains to be established whether small galaxies can
commonly host nuclear black holes and, if so, whether the same mass
versus dispersion relation mentioned above is obeyed.

Motivated by these and similar issues, \citet{gre04} conducted the
first systematic search for a population of black holes with mass
below about $10^6~M_\odot$.  They used SDSS DR1 to select a sample of
broad-lined active galactic nuclei (AGNs) with virial mass estimates
$10^{5-6}~M_\odot$.  This sample of 19 objects was the first uniformly
selected sample of low-mass AGNs.  The sample is characterized by
Eddington ratios, $L_{\rm bol}/L_{\rm Edd}$, that are close to unity,
implying radiatively efficient accretion onto these black holes.  With
its low-mass black holes and high Eddington ratios, this sample probes
a poorly explored region of parameter space, potentially providing new
insights into the physical drivers of radio properties in AGNs.  For
this reason, \citet{gre06} observed the sample at 4.9 GHz with the
Very Large Array (VLA).  A single object, \object{GH 10}, was detected
with a 4.9-GHz--to--optical flux ratio $R \sim 2.8$, formally
radio-quiet \citep{kel89}.  A stacked image of the remaining 18
objects yielded an upper limit of $R \le 0.27$, emphasizing the
atypical traits of GH\,10.

To help understand the physical origin of these atypical traits, new
multi-frequency VLA data were obtained on GH\,10 and are presented in
Section 2.  As summarized in Section 3.1, the radio emssion from
GH\,10 is found to be compact, steep-spectrum, unpolarized, and steady
in time.  A star-formation origin for this emission is considered, and
dismissed, in Section 3.2, while Section 3.3 explores the implications
of a black-hole origin for the radio emission.  The paper ends, in
Section 4, with a summary of the key findings and suggestions for
future directions.

The following cosmological parameters are assumed: $H_0 =
71$~km~s$^{-1}$~Mpc$^{-1}$, $\Omega_m = 0.27$, and $\Omega_\Lambda =
0.75$ \citep{spe03}, leading to a luminosity distance of 363~Mpc
\citep{gre04,gre06}.  Also, the sign convention adopted for the
spectral index $\alpha$ near a frequency $\nu$ is the flux density
$S_\nu \propto \nu^{+\alpha}$.

\section{VLA Imaging}

The VLA was used in the A configuration \citep{tho80} to observe
GH\,10 under proposal code AG745 on UT 2007 Jun 18 and 2007 Jul 15.
Observations were phase-referenced to the calibrator J1224+0330 whose
positional accuracy was 0.01\arcsec.  The switching angle was
5\arcdeg, while the switching times were 560~s, 260~s, and 200~s for
observations at center frequencies of 1.4000, 4.8601, and 8.4601~GHz,
respectively.  These frequencies will be abbreviated as 1.4, 4.9, and
8.5~GHz hereafter.  The {\em a priori\/} pointing position for GH\,10
was centered 3\arcsec\, south of the SDSS position \citep{gre04} to
avoid any phase-center artifacts.  Observations were made assuming a
coordinate equinox of 2000.  Data were acquired with a bandwidth of
100~MHz for each circular polarization.  Observations of 3C\,286 were
used to set the amplitude scale to an accuracy of about 3\%.

On 2007 Jun 18, net exposure times were 1500~s, 7800~s, and 8300~s at
1.4, 4.9, and 8.5~GHz, respectively.  Twenty-five to 27 antennas
provided data of acceptable quality, with most of the data loss
attributable to EVLA retrofitting activities.  The observations at
1.4~GHz were made at a large hour angle and were followed by
alternating observations at 4.9 and 8.5~GHz to sample similar hour
angles.  On 2007 Jul 15, 1.4-GHz observations were made at transit and
resulted in a net exposure time of 2800~s.  Twenty-four antennas
provided data of acceptable quality, with most of the data loss
attributable to EVLA retrofitting activities and software failures.
The data were calibrated using the 2007 Dec 31 release of the NRAO
AIPS software.  No self-calibrations were performed.  No polarization
calibration was performed, as only upper limits to the linear
polarization percentages were sought.  The AIPS task {\tt imagr} was
used to form and deconvolve images of the emission from GH\,10.

Following \citet{wro00}, {\tt imagr} was used to form and deconvolve
Stokes $I\/$ images at 4.9 and 8.5~GHz at a matched resolution at FWHM
of 0.42\arcsec.  This led to detections of a compact source at both
frequencies, with no evidence for adjacent emission on larger scales.
The 4.9~GHz image appears in Figure~1.  Quadratic fits in the image
plane yielded the integrated flux densities and the two-point spectral
index given in Table~1.  These flux densities from 2007 Jun 18 are
plotted in Figure~2.  The image fits also yielded identical
deconvolved diameters of less than 0.21\arcsec, as well as almost
identical positions.  The 4.9-GHz position is $\alpha(J2000) = 12^{h}
40^{m} 35^{s}.825$ and $\delta(J2000) = -00^{\circ} 29' 19''.53$, with
a 1-D error of 0.1\arcsec, the quadratic sum of terms due to the
signal-to-noise ratio (SNR) of the detection (less than 0.01\arcsec),
the phase-calibrator position error (0.01\arcsec), and the
phase-referencing strategies (estimated to be 0.1\arcsec).  This
4.9-GHz position is consistent with the SDSS position \citep{gre04}
marked in Figure~1.  {\tt imagr} was also used to form
matched-resolution images of Stokes $Q\/$ and $U\/$ at 4.9 and
8.5~GHz.  Those images led to no detections, with the most
constraining upper limit to the linear polarization percentage being
less than 11\% at 4.9~GHz.

Following \citet{ho01b}, {\tt imagr} was used in a two-step approach
to form and deconvolve Stokes $I\/$ images at 1.4~GHz.  First, a
search for confusing sources was done by using the inner array
baselines to image, at low resolution, the primary antenna beam to the
half-power point.  Second, all array baselines were used to image
GH\,10 and its confusing sources, simultaneously and at a high
resolution of about 1.4\arcsec\, at FWHM.  From the GH\,10 data
acquired at transit on 2007 Jul 15, this approach led to a
high-resolution image mildly affected by confusion, showing a compact
detection plus no evidence for adjacent emission on larger scales; a
quadratic fit in the image plane yielded the integrated flux density
in Table~1 and a deconvolved diameter of less than 0.72\arcsec.  From
the GH\,10 data acquired at a large hour angle on 2007 Jun 18, the
high-resolution image was more degraded by confusion so only a
parabolic fit in the image plane was made.  This yielded the peak flux
density in Table~1 that is plotted in Figure~2.  That figure also
shows the spectral index $\alpha = -0.76\pm0.05$ derived from all the
photometry on 2007 Jun 18.

Finally, the 4.9-GHz data were used to produce an image at the FWHM
resolution of the 1.4-GHz data on 2007 Jun 18.  A parabolic fit to
that 4.9-GHz image gave the flux density listed in Table~1, leading in
turn to the tabulated spectral index between 1.4 and 4.9~GHz.  Also,
that 4.9-GHz flux density is consistent with the value obtained at a
FWHM resolution of 0.42\arcsec\, implying no additional emission on
scales between 0.42\arcsec and about 1.4\arcsec.

\section{Implications}

\subsection{Radio Properties}

The new VLA data imply that the emission from GH\,10 is compact, with
a diameter of less than 0.21\arcsec\, (320~pc) at 4.9~GHz (Fig.~1) and
8.5~GHz, and less than 0.72\arcsec\, (1.1~kpc) at 1.4~GHz.  These
upper limits are derived from data with SNRs of 15 or more.  The new
1.4-GHz images show only a compact component and do not confirm the
prior hint from low-SNR data of marginally resolved emission with a
FWHM diameter of 5\arcsec\, \citep[7.5~kpc;][]{whi97}.  Similarly, the
new 4.9-GHz image does not confirm the prior suggestion from low
SNR-data of marginally-resolved emission with a FWHM diameter of
0.3\arcsec\, \citep[530~pc;][]{gre06}.

The emission from GH\,10 has a steep radio spectrum, with a
three-point spectral index of $\alpha = -0.76\pm0.05$ measured on 2007
Jun 18 and thus free from concerns about time variability (Fig.~2).
This spectral index agrees with the two-point, matched-resolution
spectral indices given in Table~1, as is expected for compact,
unresolved emission.  The single-epoch radio spectrum shown for GH\,10
in Figure~2 - the first ever measured for any of the 19 low-mass AGNs
discovered by Greene and Ho \citep{gre04,gre06} - is consistent with
optically thin synchrotron emission.  This emission is less than 11\%
linearly polarized at 4.9~GHz, well below the 73\% expected for
electrons with a power-law energy distribution, with index $p = 1 -
2\alpha$, in a uniform magnetic field \citep[][page 197]{shu91}.
Faraday depth effects could also be at play.

Variability timescales for AGNs are expected to scale with black-hole
mass \citep[e.g., ][]{ede99}, and have not yet been explored for any
of the 19 low-mass Greene and Ho objects.  The new VLA photometry can
be used to assess GH\,10's time variabilty, for future comparison with
other GH\,10 timescales to search for physical linkages.  Using only
A-configuration data, the ratio of the 4.9-GHz flux density on 2007
Jun 18 (0.75$\pm$0.05~mJy) to that on 2004 Oct 9
\citep[0.7$\pm$0.1~mJy;][]{gre06} is 7$\pm$16\%, consistent with
steady emission on timescales of years.  The ratio of the 1.4-GHz flux
density on 2007 Jul 15 (1.88$\pm$0.13~mJy) to that on 2007 Jun 18
(1.80$\pm$0.09~mJy) is 4$\pm$9\%, also consistent with steady emission
on timescales of weeks.  Also, folding in the B-configuration survey
of \citet{whi97}, the 1.4-GHz flux density of GH\,10 on 1998 Aug 11
was 1.16$\pm$0.16 mJy (peak) and 1.29$\pm$0.16 mJy (integrated).  For
such a low-SNR and compact source, \citet{dev04} advocate using the
peak flux density and seeking variability at the 4 $\sigma$, or
greater, level.  For GH\,10, the ratio of the flux density on 2007 Jul
15 (1.88$\pm$0.13 mJy) to that earlier peak (1.16$\pm$0.16 mJy) is
62$\pm$16\%, a change of less than 4 $\sigma$ on a timescale of almost
a decade.

In summary, the radio emssion from GH\,10 is compact, steep-spectrum,
unpolarized, and steady - although sparsely sampled - on timescales of
weeks and years.  The integrated 1.4-GHz flux density corresponds to a
power of $3.0 \times 10^{22}$~W~Hz$^{-1}$, while that at 4.9~GHz
corresponds to a power of $1.2 \times 10^{22}$~W~Hz$^{-1}$.

\subsection{Radio Emission from Star Formation?}

The radio emission from GH\,10 might be driven, in part, by
circumnuclear star formation.  Applying {\tt scanpi} to {\em Infrared
Astronomical Satellite\/} data at 60~$\mu$m, GH\,10 is undetected with
a 3 $\sigma$ upper limit of 0.15~Jy.  Equation (2) of \citet{yun01}
yields an upper limit to the 60-$\mu$m power.  Inserting that limit
into the radio-infrared relation in equation (4) of \citet{yun01} sets
an upper limit of $< 1.9 \times 10^{22}$~W~Hz$^{-1}$ to the 1.4-GHz
power expected from any star formation.  This limit is below the
observed power of $3.0 \times 10^{22}$~W~Hz$^{-1}$, but not by much.

A more telling analysis follows from consideration of the host
galaxy's properties, especially as they relate to estimates for a
star-formation rate.  Keck spectroscopy, with a 1\arcsec\, (1.5~kpc)
aperture, exhibit the Mg I{\it b} and Ca II infrared triplets
characteristic of an old stellar population \citep{bar05}.  {\em
Hubble Space Telescope\/} $B$ and $I$ imaging, with a resolution at
FWHM of 0.12\arcsec\, (170~pc), reveals a smooth, spheroidal-looking
system with a steep Sersic index $n=4$ and a color $B-I=2$ that is
characteristic of Sa or spheroidal galaxies \citep{gre08}.  SDSS
spectroscopy with a 3\arcsec\, (4.5~kpc) aperture shows narrow
H$\alpha$ and [O~II] $\lambda$3727 lines but, as a Balmer decrement
could not be robustly measured, an estimate of the internal reddening
is unavailable for GH\,10 \citep{gre04}.  However, if the average
reddening advocated for the 19 Greene and Ho objects by \citet{gre08}
is adopted and the reddening curve of \citet{car89} is applied, then
the correction factors for narrow H$\alpha$ and [O~II] $\lambda$3727
are 1.3 and 1.7, respectively.  Then using equations (2)-(3) of
\citet{ken98}, those corrected line luminosities for GH\,10 each imply
a star formation rate no higher than 2~$M_\odot$~yr$^{-1}$.  These
rate estimates are, by construction, upper limits because they neglect
the portion of the line-emitting gas that is energized by the AGN
rather than by star formation.  Moreover, in order for GH\,10 to be
within the 2 $\sigma$ range of the radio-infrared relation mentioned
above, the reddening at H$\alpha$ would need to be 2 mag, far more
than the average (0.3 mag) and largest (0.5 mag) reddening advocated
for the Greene and Ho objects \citep{gre08}.  Fortunately, GH\,10 is
also detected in a {\em Spitzer\/} IRS spectrum, boding well for a
future estimate of a star formation rate that is less affected by
reddening concerns (C. E. Thornton et al., in preparation).

For comparison, using equation (13) of \citet{yun01}, the observed
1.4-GHz power of GH\,10 would correspond to a star-formation rate of
18~$M_\odot$~yr$^{-1}$.  Such a high rate is clearly inconsistent with
the line-based upper limit of 2~$M_\odot$~yr$^{-1}$ derived above.
Circumnuclear star formation cannot, therefore, dominate the radio
emission from GH\,10.

\subsection{Radio Emission from the AGN}

Based on the above analysis, it seems plausible that the radio
emission from GH\,10 is mainly energized by its low-mass black hole.
GH\,10 is radiating at about twice its Eddingtion luminosity, with
this estimate being uncertain by a factor of 3 \citep{gre04}.  Thus
comparisons of GH\,10 with other broad-lined AGNs with near-unity
Eddington ratios are especially apt.  The 4.9-GHz--to--optical flux
ratio for GH\,10 implies $R \sim 2.8$, so GH\,10 is radio-quiet
\citep{kel89}, a trait shared by most quasars at $z < 0.5$
\citep{ho01a,gre04,gre06}.  There is also a faint, compact {\em
Chandra\/} detection within 1\arcsec\, (1.5~kpc) of the SDSS position
of GH\,10, and its spectral properties resemble those of high-mass
black holes \citep{gre07a}.  The ratio of the 4.9-GHz luminosity, $5.9
\times 10^{38}$~ergs~s$^{-1}$, to the hard X-ray luminosity, $1.9
\times 10^{42}$~ergs~s$^{-1}$, implies $R_{\rm X} = 10^{-3.5}$
\citep{ter03}.  Such a value is also consistent with the values shown
by most $z < 0.5$ quasars and by most luminous Seyfert 1 galaxies
\citep{ho08}.  Unfortunately, such systems are too faint and too
distant to be resolved on sub-kiloparsec scales in VLA images, so no
structural comparisons can be made.

GH\,10's values for $R\/$ and $R_{\rm X}$, plus its radio and hard
X-ray luminosities, also resemble those for the more luminous Palomar
Seyfert galaxies \citep{pan07}.  VLA imaging at 1.4 and 4.9~GHz of the
Palomar Seyfert galaxies, with a resolution of about 1\arcsec\,
(100~pc at the median distance of 20~Mpc), shows a wide range of radio
powers ($10^{18}-10^{25}$~W~Hz$^{-1}$), spectral indices (+0.5 to -1),
and linear sizes (a few tens of parsecs to 15~kpc)
\citep{ho01b,ulv01}.  Their structures mainly show a compact core,
either unresolved or slightly resolved, sometimes with adjacent
elongated, jetlike/outflow components.  Linearly polarized emission,
while rarely detected, is preferentially associated with features
adjacent to the cores.  Thus GH\,10's radio properties most resemble
those of the steep-spectrum cores of the Palomar Seyferts, for which
the emission is unmodified by the effects of internal and/or external
opacity.  Like the Palomar Seyferts, the steep-spectrum emission from
GH\,10 could be outflow-driven.  Moreover, the lower escape velocity
from GH\,10's dwarf host galaxy \citep{bar05} would make it easier for
any synchrotron-emitting plasma to leave the vicinity of the AGN,
thereby favoring a steep, rather than flat or inverted, radio
spectrum.

This work has established that the radio emission from GH\,10 is
compact, steep-spectrum, unpolarized, and steady.  Still, GH\,10 is
known to be atypical when compared to the other GH objects, which are
especially radio-quiet with $R \le 0.27$ from a stacking analysis
\citep{gre06}.  Following those authors, by analogy with supermassive
and stellar-mass black holes, a near-unity Eddington ratio suggests
that the low-mass black hole GH\,10 is presently in the very high
state, with its entry into that state from the high/soft state
accompanied by an optically thin ejection event in the radio.  Within
this framework, and barring opacity effects, the radio emission from
GH\,10 would be the slowly fading, steep-spectrum remnant from that
ejection event, whose power corresponds to the quenched radio levels
expected for the high/soft state \citep[e.g.,][]{mac03}.  Also, a
radio detection of only one of the 19 GH objects suggests, at face
value, that such radio ejections from GH objects occur only about 5\%
of the time.  Interestingly, a 5\% duty cyle resembles that
established observationally for radio flares from microquasars
\citep[e.g.,][]{nip05}.

Because the low-mass black hole GH\,10 is radiating close to its
Eddington limit, it may be a local analog of the starting conditions,
or seeds, for supermassive black holes.  From the present work, the
diameter of GH\,10's radio emission must be less than 0.21\arcsec\,
(320~pc).  If GH\,10 is indeed driving some of this emission, then
1.4-GHz imaging with Very Long Baseline Interferometry (VLBI) at a
resolution of 0.005\arcsec\, (7.5~pc) will reveal the physical
connection between the black hole and the synchrotron-emitting plasma.
Although GH\,10 is a faint radio source (Table~1), VLBI imaging of
such sources is feasible \citep[e.g.,][]{wro06} and would offer a
means to study the relative roles of radiative versus kinetic feedback
during black-hole growth \citep[e.g.,][]{dim05,kuh05}.

\section{Summary and Future Directions}

GH\,10 was the only object detected by \citet{gre06} in their VLA
survey of the 19 low-mass AGNs discovered by \citet{gre04}.  New VLA
imaging reveals that the emssion from GH\,10 has an extent of less
than 320~pc, with an optically thin synchrotron spectrum that is less
than 11\% linearly polarized.  The radio emission is steady, although
poorly sampled, on timescales of weeks and years.  Circumnuclear star
formation cannot dominate the radio emission, because the inferred
star formation rate, 18~$M_\odot$~yr$^{-1}$, is inconsistent with the
upper limit of 2~$M_\odot$~yr$^{-1}$ derived from the narrow optical
emission lines.  Instead, the radio emission must be dominantly
energized by the low-mass black hole, which is thought to be radiating
near its Eddington limit.

The radio properties of GH\,10 strongly resemble those of the
steep-spectrum cores of Palomar Seyfert galaxies, which feature
outflow-driven emission unmodified by the effects of internal and/or
external opacity.  This suggest that, like those Palomar Seyferts, the
radio emission from GH\,10 is outflow-driven, a suggestion that can be
tested with future VLBI imaging.  Such imaging could also offer an
opportunity to study the relative roles of radiative versus kinetic
feedback during black-hole growth, given that GH\,10 may be a local
analog of the starting conditions, or seeds, for supermassive black
holes.  Finally, the compact, steep-spectrum, unpolarized, and steady
nature of the emission from GH\,10 will guide future observations of
the new 1.4-GHz detections of low-mass AGNs reported by
\citet{gre07b}.


\acknowledgments We acknowledge useful feedback from an anonymous referee.

{\it Facilities:} \facility{VLA}

\clearpage

\begin{deluxetable}{ccclc}
\tablecolumns{5}
\tablewidth{0pc}
\tablecaption{VLA Photometry of GH\,10}\label{tab1}
\tablehead{
\colhead{}      & \colhead{Resolution} &
\colhead{$\nu$} & \colhead{$S_\nu$}    & \colhead{} \\
\colhead{Date}  & \colhead{(\arcsec,\arcsec,\arcdeg)} &
\colhead{(GHz)} & \colhead{(mJy)}      & \colhead{Fig.~2} \\
\colhead{(1)}   & \colhead{(2)}        &
\colhead{(3)}   & \colhead{(4)}        & \colhead{(5)}}
\startdata
2007 Jun 18& 0.42,0.42,0  & 4.9& 0.75$\pm$0.05\tablenotemark{a}& Yes \\
           & 0.42,0.42,0  & 8.5& 0.44$\pm$0.03\tablenotemark{a}& Yes \\
           & 1.96,1.44,-47& 1.4& 1.80$\pm$0.09\tablenotemark{b}& Yes \\
           & 1.96,1.44,-47& 4.9& 0.72$\pm$0.04\tablenotemark{b}& No  \\
2007 Jul 15& 1.45,1.36,47 & 1.4& 1.88$\pm$0.13                 & No  \\
\enddata
\tablecomments{
Col.~(1): UT observation date.
Col.~(2): Elliptical Gaussian resolution at FWHM and its elongation
position angle.
Col.~(3): Center frequency.
Col.~(4): Flux density with error given by the quadratic sum of the
3\% scale error and the 1 $\sigma$ rms of the fit.
Col.~(5): Plotted in Figure 2?}
\tablenotetext{a}{Two-point spectral index $\alpha = -0.97\pm0.17$.}
\tablenotetext{b}{Two-point spectral index $\alpha = -0.73\pm0.06$.}
\end{deluxetable}

\clearpage

\begin{figure}
\epsscale{1.0}
\plotone{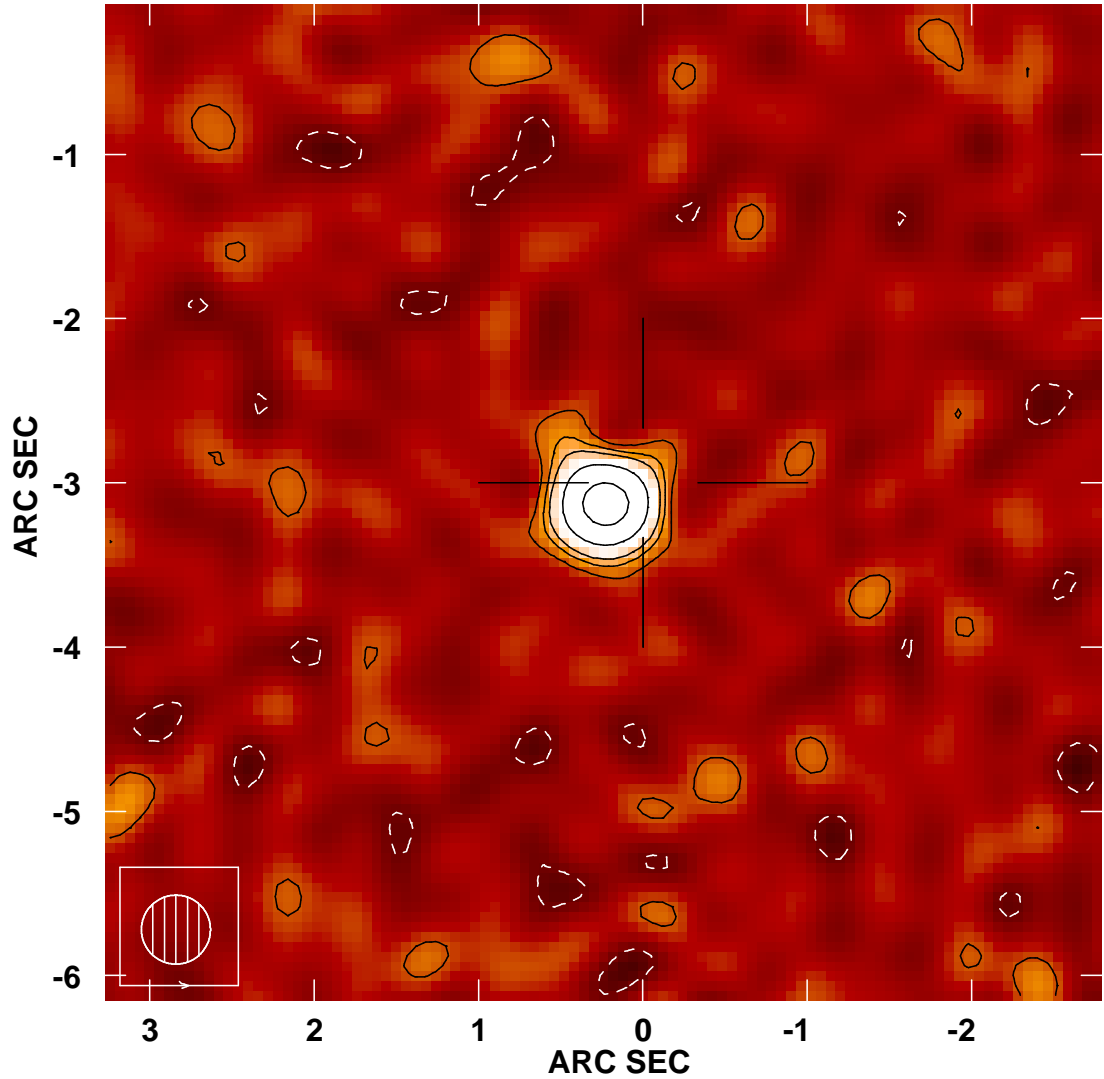}
\caption{VLA image of Stokes $I\/$ emission from GH\,10 at a frequency
of 4.9~GHz and spanning 6\,arcsec\, (9~kpc).  Coordinate labels are
relative to the pointing position, which was centered 3\arcsec\, south
of the SDSS position to avoid phase-center artifacts.  SDSS position
is marked with a cross with a gap.  Natural weighting was used, giving
an rms noise of 0.022~mJy~beam$^{-1}$ (1 $\sigma$) and Gaussian beam
dimensions at FWHM of 0.42\arcsec\, (hatched circle).  Contours are at
$-$6, $-$4, $-$2, 2, 4, 6, 12, and 24 times 1 $\sigma$.  Negative
contours are dashed and positive ones are solid.  Linear color scale
spans $-$0.1~mJy~beam$^{-1}$ to 0.2~mJy~beam$^{-1}$.}\label{fig1}
\end{figure}
\clearpage

\begin{figure}
\epsscale{1.0}
\plotone{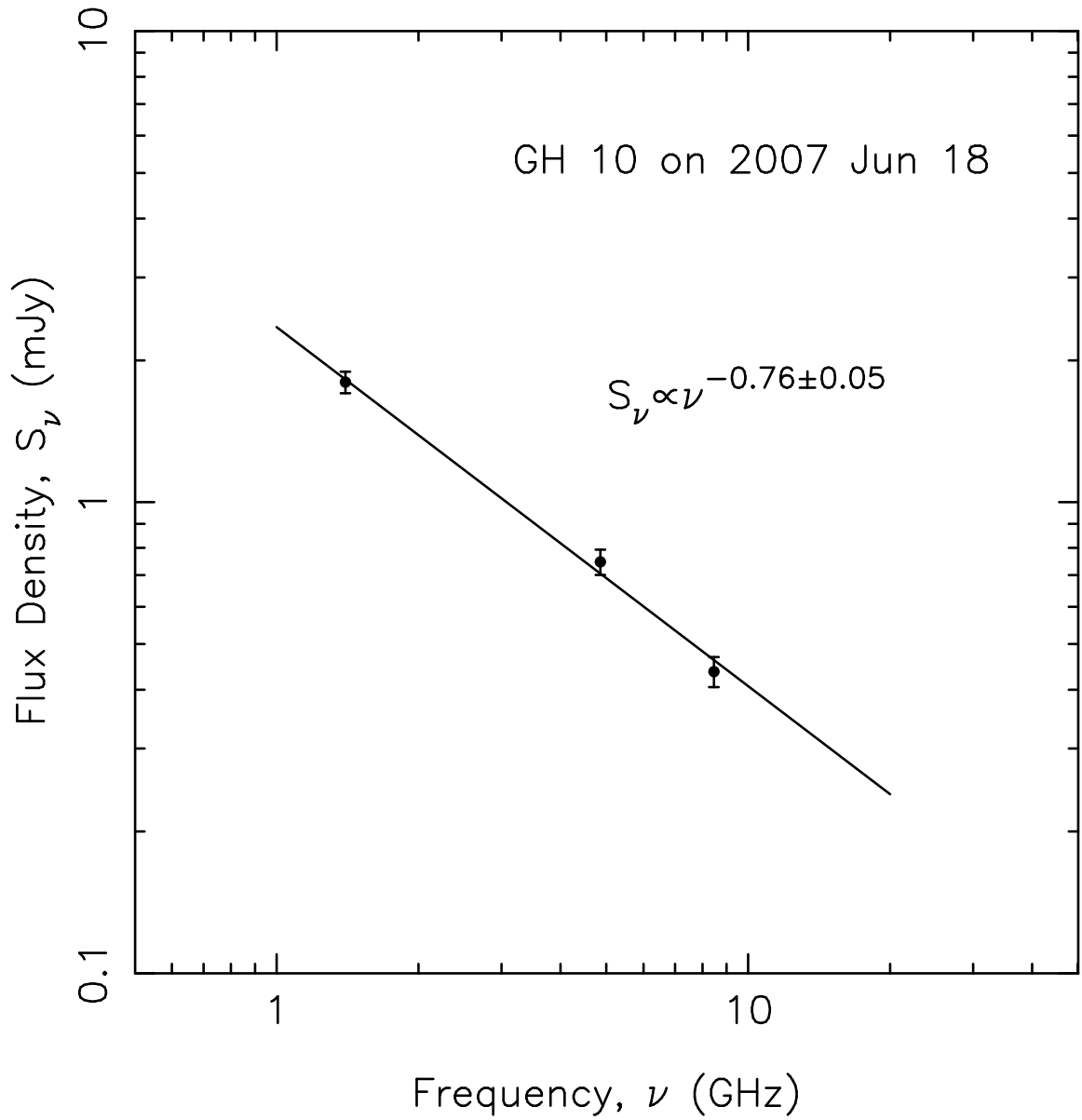}
\caption{VLA spectrum of GH\,10 as measured at 1.4, 4.9, and 8.5~GHz
on 2007 Jun 18.}\label{fig2}
\end{figure}

\end{document}